\documentclass[twocolumn,showpacs,preprintnumbers,pre,floatfix]{revtex4}
\usepackage{amssymb,amsmath,amsfonts}
\usepackage{graphicx,epsfig} % Include figure files
\usepackage{bm} % bold math

\begin{document}
\title{Small-world behavior in time-varying graphs}
\author{J. Tang,$^{1}$ S. Scellato,$^{1}$ M. Musolesi,$^{1}$ C. Mascolo,$^{1}$ V. Latora,$^{2,3}$}
\affiliation{%
$^1$ Computer Laboratory, University of Cambridge, 15 JJ Thomson Av.,  
Cambridge CB3 0FD, UK}
\affiliation{%
$^2$~Dipartimento di Fisica e Astronomia, Universit\`a di Catania and INFN, 
Via S. Sofia, 64, 95123 Catania, Italy}
\affiliation{%
$^3$~Laboratorio sui Sistemi Complessi, Scuola Superiore di Catania, 
Via San Nullo 5/i, 95123 Catania, Italy}
\date{\today}
\begin{abstract}
  Connections in complex networks are inherently fluctuating over time
  and exhibit more dimensionality than analysis based on standard
  static graph measures can capture. Here, we introduce the concepts
  of temporal paths and distance in time-varying graphs. We define as
  temporal small world a time-varying graph in which the links are
  highly clustered in time, yet the nodes are at small average
  temporal distances.  We explore the small-world behavior in
  synthetic time-varying networks of mobile agents, and in real social
  and biological time-varying systems.
\end{abstract}

\maketitle %

In the last decade, the study of complex networks has attracted a lot
of attention in the scientific community as various social, biological
and technological systems can be represented and analysed as graphs
\cite{reviews}.  Typically, such systems are inherently dynamic, with
the links changing and fluctuating over time.  Human contacts or
relationships change over time because individuals lose old
acquaintances, acquire new ones, or move over geographic space
\cite{gonzalez,buscarino}. Communication in man-made networks, such as
machine connections and social interactions over the Internet, takes
place at specific points in time \cite{holme_network_2005,
  kossinets_structure_2008, tadic_04}. New links appear while some
others disappear in the World-Wide-Web \cite{tadic_01}, in patterns of
interactions among gene from microarray time experiments
\cite{caretta07,ahmed} or in functional brain networks
\cite{valencia,fallani}.  The time evolution of a network by the
addition, as well as the deletion, of nodes and links has been
extensively modelled with the main purpose of reproducing,
asymptotically, statistical properties such as scale-free degree
distributions \cite{reviews,dorogotsev01}.  With only a few notable
exceptions \cite{gautreau,raval,stehle}, less focus has been given to
the characterization of the dynamics of complex networks in stationary
conditions.

In particular, the small-world phenomenon, i.e., the fact that real
networks have high clustering coefficient, while the typical distance
between their nodes is small as in random graphs, has been
investigated in {\em static graphs}, neglecting the temporal dimension
\cite{watts_sw_98,latora_efficient_01,delosrios}.  The time evolution
of a real system, when considered, is usually studied by evaluating
the standard static measures (distances and clustering coefficient) on
snapshots of the network taken at different times
\cite{vicsek,clauset_persistence_2007}.  As we will show below, this
approach does not capture entirely the dynamic correlations of a 
time-varying network. In this Letter, we introduce 
a measure of the temporal distance between the nodes of  {\em
  time-varying graphs}, i.e. graphs with time-fluctuating links. Such
temporal distance takes into account the actual time order, duration
and correlations between links appearing at different times. This metric, 
together with a measure of the time-persistence of the links, 
allows to define and investigate temporal
small-world behavior in social and biological networks that change
over time.

Consider a network with $N$ nodes, where the links
can fluctuate in time. The typical example is a social system,
with no births or deaths, where the patterns of interaction are
changing in time because of the spatial movement of the individuals,
or because the individuals lose old acquaintances and get new ones.
The system can be described at its maximum resolution sampling time as
a {\em time-varying graph}, i.e. a discrete sequence (an ordered set)
$G_1,G_2,\ldots,G_T$ of $T$ undirected or directed graphs, where $T$
is the length of the sequence.  In compact notation, we denote the
entire sequence as ${\cal G}=\{G_t\}_{t=1,2,\ldots,T}$. A time-varying
graph ${\cal G}$ can be represented by means of a $N \times N$
time-dependent adjacency matrix $A(t), ~t=1,\ldots,T$, where
$a_{ij}(t)$ are the elements of the adjacency matrix of the $t$-th
graph. We indicate as $K(t)$ the number of links in the $t$-th graph
of the sequence. A sequence of graphs is convenient to describe
systems where each connection starts at a specific time, and also has
a temporal duration. In this sense, time-varying graphs are different
from previous temporal approaches \cite{holme_network_2005,
  kossinets_structure_2008,
  kempe_connectivity_2002,kostakos_temporal_2009} designed to
characterize systems as email exchanges, where the links have instead
no temporal duration, because the exchange is instantaneous.
Moreover, a time-varying graph is a different ensemble from those
usually studied in the literature \cite{park,bianconi}.  In fact, in a
time-varying graph, what matters is not only the probabilty
distribution $P(G)$ over the graphs in the ensemble, but also how the
graphs are ordered in time.  By counting the number of times a given
graph $G$ appears in the time sequence, we can construct $P(G)$. To
fully describe time-varying graphs we also need to know how graphs are
correlated in time. For instance we need to know the conditional
probabilities $P(G_t|G_{t-1})$ of observing graph $G_t$ after graph
$G_{t-1}$ (more in general, the probabilities
$P(G_t|G_1,G_2,\ldots,G_{t-1})$ of observing graph $G_t$ after the
sequence $G_1,G_2,\ldots,G_{t-1}$). In most cases, the contacts
between the same node pair in time-varying systems tend to be
clustered in time, i.e. they show persistence over time
\cite{holme_network_2005}.  For instance, people tend to engage
in relations for continuous intervals of time.  Hence, a given link
has a higher probability to appear in graph $G_t$ if it was already
present in graph $G_{t-1}$. To quantify this effect, following
Ref.~\cite{clauset_persistence_2007} we compute $C$, the average
topological overlap of the neighbor set of a node between two
successive graphs in the sequence:
\begin{equation}
C = \frac{\sum_i C_i}{N}
~~~ C_i = \frac{1}{T-1} \sum_{t=1}^{T-1}\frac{ \sum_j a_{ij}(t) a_{ij}(t+1) }
           {      \sqrt { [ \sum_j a_{ij}(t) ]     [ \sum_j a_{ij}(t+1) ] }           }
\end{equation}
We name this metric the {\em temporal-correlation coefficient} of $\cal G$.
The value of C is in the range [0,1]. In particular, if 
all graphs in the sequence are equal, we have $C=1$.

\begin{figure}[!htbp]
   \centering
\includegraphics[width=7.8cm,angle=0]{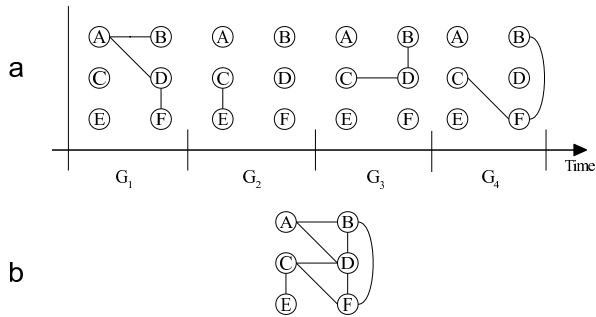}
\caption{An example of a time-varying graph $\cal G$ with $T=4$ (panel a), and its
  projection into a static graph (panel b).}
\label{fig04:example}
\end{figure}

A fundamental concept in graph theory is that of geodesic, or shortest path.   
In a static graph, a shortest path between 
nodes $i$ and $j$ is defined as a path of minimal length between the two 
nodes. This is a sequence of adjacent nodes starting at $i$, ending at $j$,  
and visiting the minimum number of nodes. Finally,  
the distance between node $i$ and node $j$ is set equal to the length
of the shortest paths from $i$ to $j$. Here, we introduce the concepts of {\em
  temporal shortest path} and {\em temporal distance} to generalize
the definitions of shortest paths and of node distance to the case
of time-varying graphs. We illustrate the basic idea with the example 
shown in Fig.~\ref{fig04:example}a. Suppose 
node $A$ wants to send a message in the fastest possible way to the
other nodes of the graph.  We assume that node $A$ can start passing the
message at time $t=1$, and
the message has to be delivered by time $t=4$. On graph $G_1$, node
$A$ can directly pass the message to nodes $B$ and $D$, which are 
therefore assigned temporal distance 1 from node $A$, since they can be
reached in one unit of time.  There are also other temporal paths to
go from $A$ to nodes $B$ and $D$ in three time units. For example, we can go
from $A$ to $D$ in the following way: $A \rightarrow B$ in $G_1$, $B
\rightarrow D$ in $G_3$. This is also a temporal path from $A$ to $D$,
though it is not the shortest, since the fastest way to go
from $A$ to $D$ is to use the link $A \rightarrow D$ in $G_1$.  
Distance 3 is assigned to node $C$, since the message can be passed from $A$
to $D$ in graph $G_1$, and then from node $D$ to node $C$ in $G_3$,
thus reaching $C$ in three time units.
Node $F$ can be reached in 4 time steps by means of three alternative shortest
paths: $A \rightarrow B$ in $G_1$, $B \rightarrow F$ in $G_4$; $A
\rightarrow D$ in $G_1$, $D \rightarrow C$ in $G_3$, $C \rightarrow F$
in $G_4$; and $A \rightarrow D$ in $G_1$, $D \rightarrow B$ in $G_3$,
$B \rightarrow F$ in $G_4$. Finally, there are no temporal paths from
$A$ to $E$, hence we set the temporal distance of $E$ from $A$ equal
to $\infty$, and we say that $E$ is not reachable from $A$. This is an
effect of the time order of the links in a time-varying graph, and indeed 
node $A$ and $E$ are connected in the aggregate graph shown in
Fig.~\ref{fig04:example}b in which all links are considered as
concurrent. 

Notice also that, due to the time order of the links, the
temporal distances are not symmetric, even if the time-varying graph
consists of a sequence of undirected graphs. For instance, while the temporal
distance from $A$ to $F$ is 4, the temporal distance from $F$ to $A$
is $\infty$ (because the links occur in the wrong time order to
facilitate the passage from $F$ to $A$). Conversely, in the static graph
in Fig.~\ref{fig04:example}b, we have $d_{AF}= d_{FA}=2$. In order words, 
the main difference between a time-varying graph {\cal G}, as that shown in
panel a), and its associated static graph, reported in panel b), is that
some of the shortest paths of the static graph are not temporally
valid (in the sense that the links do not appear in the correct time
order) and, therefore, cannot be used to route messages. 
In general, in time-varying graphs there are
more disconnected node pairs, than in static ones. As an example,
the static graph in Fig.~\ref{fig04:example} is composed of a single
connected component, while if  time is taken into consideration, it 
is not possible to go from $A$ to $E$, or from $F$ to $A$.  In order to compute the temporal distances $d_{ij}$ 
for all node pairs $i,j=1,2,...,N$ of a generic graph $\cal G$,  
we have implemented a generalization of the breadth
first search algorithm. 
The average temporal connectivity properties of $\cal G$ can be measured by the 
{\em characteristic temporal path length} $L$:  
\begin{equation}
  L =   \frac{1}{N(N-1)} \sum_{ij} d_{ij}  
\end{equation}   

Alternatively, in order to avoid the potential divergence due to pairs
of nodes that are not temporally connected, we can define the {\em
  temporal global efficiency} of $\cal G$ as
\cite{latora_efficient_01}:
\begin{equation}
E =   \frac{1}{N(N-1)} \sum_{ij} \frac{1}{d_{ij}}  
\end{equation}  
Low values of $L$ (high values of $E$) indicate that the nodes of the
graphs can communicate efficiently. In the following, we will show
that time-varying graphs from models and real-world systems can be, 
at the same time, temporally clustered and still have small temporal
distances between their nodes. In analogy with the
small-world analysis in static graphs \cite{watts_sw_98,latora_efficient_01}, 
we will compare the actual values of $C$, $L$ and $E$ of a given time-varying graph $\cal G$,
with the corresponding values calculated by considering an ensemble
$\{ {\cal G}^{rand} \}$ of randomized versions of $\cal G$. Each
sequence $ {\cal G}^{rand}$ is obtained by randomly reshuffling the
graphs in $\cal G$, i.e., by destroying the time order (and
correlations) in the original sequence $G_1,G_2,\ldots,G_T$. More
precisely, we will show that some time-varying graphs can have a value
of $C$ much larger than the correlation coefficient of the reshuffled
sequence $C^{rand}$, and, at the same time a value of $L$ as small as
$L^{rand}$. We will refer to this behavior as {\em small-world
  behavior in time-varying systems}.
\begin{figure}[!htbp]
   \centering
\includegraphics[width=7.0cm,angle=0]{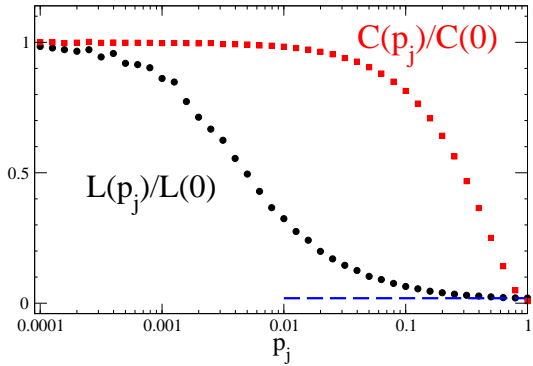}
\caption{Characteristic temporal path length and temporal-correlation
  coefficient of time-varying graphs produced by the model of moving
  agents, as a function of the probability $p_j$ of long-distance
  jumps. In the simulations we have set $N=100$, $D=100 ~m$, $v=1 ~
  m/s$, $r_c=5 ~m$ and produced sequences of length $T=500$.  The
  characteristic temporal path length of the reshuffled sequences is
  reported as dashed line.  }
\label{fig2}
\end{figure}

{\em{Random-walkers network model.-}}
We first illustrate how this behavior can be obtained in a network
model of moving agents, as a result of simple motion rules.  We
consider a system of $N$ random walkers which move in a
two-dimensional square of linear size $D$ with a fixed velocity $v$,
and additionally perform long-distance jumps to randomly chosen
position of the square with a jump probability $p_{j}$
\cite{buscarino}.  For each fixed value of $p_{j}\in [0,1]$, the
time-varying network $\cal G$ is constructed by linking, every second,
all nodes having a distance in space smaller than a given value $r_c$. 
In Fig.~\ref{fig2} we plot $C$ and $L$ as a function of $p_j$.  The
values reported are normalized to the maximum values of $C$ and $L$
obtained for $p_j=0$, and respectively equal to $C(0)=0.91$ and
$L(0)=442.8$.  We observe that a small percentage of jumps is
sufficient to create links between nodes otherwise at large temporal
distances and to produce a large drop in the temporal $L$. When
$p_j=0.01$, $L$ has reduced to one forth of $L(0)$, and when
$p_j=0.1$, $L$ has about the same value as for the reshuffled
sequence.  The value of $L^{rand}$ obtained as an average over 1000
realizations of ${\cal G}^{rand}$ is reported as dashed line.  
While $L(p_j)$ is rapidly decreasing, $C(p_j)$ is constant up 
to large values of $p_j \sim 0.1$, so that for intermediate values of $p_j$ we have 
time-varying graphs exibiting small-world behavior. 
Finally, we have found that, by approximating for each value of $p_j$ 
the corresponding time-varying graph $\cal{G}$ as a static graph, we
obtain a value of static L (not reported in figure) which slowly
changes in the interval $[0,1]$. For instance, for $p_j=1$, we have
$L(p_j)/L(0)=0.74$. We thus cannot capture the temporal small-world
behavior with the standard characteristic path length of a static
graph.
%\cite{note2}

{\em{ Brain cortical networks.-}}
We finally explore real-world time-varying complex networks. We first
consider time-varying functional cortical networks extracted from a
set of high-resolution EEG recordings in a group of 5 normal subjects
performing a task consisting in a foot movement \cite{fallani}.  For
each subject, and for each of four frequency bands ($\alpha, \beta,
\gamma, \theta$), we considered a time period of 0.5 sec corresponding
to the final phase of execution of the foot movement.  Each
time-varying graph has $N=16$ nodes, representing cortical regions of
interest, and consists in a time sequence of $T=100$ directed
unweighted graphs, where the directed links represent causal
influences between cortical regions (see Ref.~\cite{fallani} for
details).
\begin{table}[!htb]
\begin{scriptsize}
\begin{tabular}{c|cccccc}
           & $C$ & $C^{rand}$   & $L$                & $L^{rand}$        & $E$    & $E^{rand}$ \cr  
  \hline
  \hline 
%  SUBJECT 1 PHASE 4 
$\alpha$  & 0.44& 0.18 (0.03)        &    3.9   $(100\%)$   & 4.2  $(98\%)$   & 0.50    & 0.48     \cr
$\beta$   & 0.40& 0.17 (0.002)        &    6.0   $(94\%)$    & 3.6  $(92\%)$   & 0.41    & 0.45     \cr
$\gamma$  & 0.48& 0.13 (0.003)        &    12.2  $(86\%)$    & 8.7  $(89\%)$   & 0.39    & 0.37     \cr
$\delta$  & 0.44& 0.17 (0.003)         &    2.2   $(100\%)$   & 2.4  $(92\%)$   & 0.57    & 0.56     \cr 
\hline
\hline
d1  & 0.80 & 0.44 (0.01)    &   8.84   $(61\%)$   &  6.00 $(65\%)$   & 0.192   &  0.209     \cr
d2  & 0.78 & 0.35 (0.01)   &   5.04   $(87\%)$   &  4.01 $(88\%)$   & 0.293   &  0.298     \cr
d3  & 0.81 & 0.38 (0.01)   &   9.06   $(57\%)$   &  6.76 $(59\%)$   & 0.134   &  0.141     \cr
d4  & 0.83 & 0.39 (0.01)   &   21.42  $(15\%)$   &  15.55$(22\%)$   & 0.019   &  0.028     \cr
\hline
\hline
Mar & 0.044 & 0.007 (0.0002)  &   456               & 451  & 0.000183 & 0.000210 \cr
Jun & 0.046 & 0.006 (0.0002)  &   380               & 361  & 0.000047 & 0.000057 \cr
Sep & 0.046 & 0.006 (0.0002)  &   414               & 415  & 0.000058 & 0.000074 \cr
Dec & 0.049 & 0.006 (0.0002)  &   403               & 395  & 0.000047 & 0.000059 \cr 
  \hline
  \hline 
\end{tabular}
\end{scriptsize}
\caption{Temporal-correlation, characteristic temporal path length and efficiency for
  brain cortical networks (subject 1, and four band frequencies)
  \cite{fallani}, for the social interaction networks of INFOCOM'06 
  (time periods between 1pm and 2:30pm, four different days), and 
  for messages over Facebook online social network (three different months 
  of year 2007) \cite{facebook}.  
  Results are compared with those obtained for $1000$ randomized (shuffled) 
  sequences of the same length.  The values in parenthesis next to $C_{rand}$ are the respective standard deviations.
  The values in parenthesis next to $L$ and $L^{rand}$ are the percentage of pairs of nodes that are temporally connected and 
  not considered in the averages.  
}
\label{tab1}
\end{table}
We have computed the values of $C$, $L$ and $E$ for each real sequence
and for the reshuffled ones. In Table~\ref{tab1} we report the results
for one of the subjects.  For all the considered bands, the real
sequence exhibits small-world properties, having a large value of $C$
(significantly larger than $C^{rand}$) and, at the same time, a small
characteristic temporal path length (a high efficiency), comparable to
that observed in the shuffled sequence. Similar results (not reported)
were obtained for the other four subjects.

{\em{ Social interaction networks.-}}
The second real case study of our analysis is a time-varying social
network based on a dataset of contacts among participants of
INFOCOM'06, a major data communication conference which took
place in a hotel. The contacts were collected by means of
Bluetooth-enabled devices able to record interactions among people
that are in promixity \cite{infocom}.  The discovery process of new
devices was performed every 2 minutes. In Table~\ref{tab1} we report
the data for the interactions during lunchtime between 1pm and
2:30pm. This is the interval with the larger number of contacts during
a day. Each sequence is made of $T=45$ undirected unweighted graphs
with $N=78$ nodes each. The average path length and the efficiency are
similar for the original and reshuffled traces (the number in
parenthesis close to $L$ and $L^{rand}$ are the percentage of pair of
nodes being temporally connected and hence considered in the
computation of the average path length), whereas $C$ is more than
double that of $C^{rand}$.  This can be considered as an indication
of small-world behavior in these traces according to our definition.

{\em{ Online social networks.-}}
The third system we study is based on interactions over an online
social network. The original dataset contains the messages sent among
6 millions users in the London network of Facebook over one year
(March 2007 to February 2008) \cite{facebook}. We have divided the
contacts according to the months of the year and, for each month, we
have filtered out all contacts between pairs of nodes which exchange
less than 10 messages per month. This allows us to consider only the
subset of most active users, obtaining networks with about $N=100,000$
users per month.  For each month, the time varying graph is composed
by $T=720$ (or $T=744$) directed graphs, one for each hour of the
month.  As shown in Table~\ref{tab1} for four different months of the
dataset, the average temporal path length of the time-varying networks
is close to the value obtained for the reshuffled sequences. However,
the network under study is disconnected in several different
components, and only an extremely small percentage (about $10^{-6}$)
of the node couples are temporally connected.  Consequently, the
characteristic temporal path length was evaluated as an average over a
small number of node couples. A better characterization of the system
can be obtained by means of the temporal efficiency. The values of $E$
and $E^{rand}$ measured for Facebook are in general smaller than those
observed in the other two networks, this being due to the high
disconnectedness of Facebook. Nevertheless, as for the case of the
cortical networks and of INFOCOM'06, the real Facebook is almost as
efficient as its reshuffled version. Finally, also for Facebook we
observe a temporal small-world behavior: while the length of the
temporal paths of the time-varying network are not affected by the
reshuffling procedure, the temporal correlation coefficient $C$ is
about one order of magnitude larger than in the reshuffled version
$C^{rand}$.

In conclusion, our results suggest that time-varying networks, strongly
clustered in time and, at the same time, with short temporal paths
between their nodes, might be widespread in biological, social and 
man-made systems, often with important dynamical consequences \cite{note}. 
We hope that our work will stimulate further studies of 
temporal small-world behavior in real time-varying systems.

%%%%%%%%%%%%%%%%%%%   ACKNOWLEDGEMENTS   %%%%%%%%%%%%

We thank F. Babiloni and F. De Vico Fallani for making the cortical
networks available to us, and Ben Y. Zhao (UCSB) for the Facebook
traces. We acknowledge the support of EPSRC through grants
EP/C544773, EP/E012914 and EP/D077273.


\begin{thebibliography}{99}

\bibitem{reviews} R. Albert and A.-L.~Barab\'asi, Rev. Mod. Phys. {\bf
    74}, 47 (2002); S.N. Dorogovtesev, J.F.F. Mendes, {\it Evolution
    of networks},(Oxford University Press, 2003); S.\ Boccaletti, et
  al., Phys. Rep. \textbf{424}, 175 (2006).
\bibitem{gonzalez} M. C. Gonz\'alez, C.A. Hidalgo and A.-L.~Barab\'asi, Nature \textbf{453}, 779 (2008). 
\bibitem{buscarino} A. Buscarino, L. Fortuna, M. Frasca, V. Latora, Europhys. Lett.  \textbf{82}, 38002 (2008). 
\bibitem{holme_network_2005} P.~Holme, Phys. Rev. E \textbf{71}, 046119 (2005).
\bibitem{kossinets_structure_2008} G.~Kossinets, J.~Kleinberg, and D.~Watts, 
arXiv:0806.3201 (2008).
\bibitem{tadic_04}  B. Tadi\'c and S. Thurner, Physica A {\bf 332}, 566 (2004)
\bibitem{tadic_01} B. Tadi\'c, Physica A {\bf 293}, 273 (2001)
\bibitem{caretta07} C. Caretta-Cartozo, P. De Los Rios, F. Piazza, P. Li\`o, PLoS Computational Biology,  \textbf{3}, e103 (2007).  
\bibitem{ahmed} A. Ahmed and E. P. Xing, Proc. Natl. Acad. Sci. USA {\bf 106}, 11878 (2009).  
\bibitem{valencia} M.\ Valencia, J.\ Martinerie, S.\ Dupont, and M.\ Chavez, Phys. Rev. E.  \textbf{77}, 050905R (2008).  
\bibitem{fallani} F. De Vico Fallani et al., Journ. Phys. A: Math. Theor.  \textbf{41}, 224014 (2008) 

\bibitem{dorogotsev01} S.N. Dorogovtsev and J.F.F. Mendes, Phys. Rev. E 
\textbf{63}, 056125 (2001).
\bibitem{gautreau} A. Gautreau, A. Barrat, and M. Barthelemy, 
Proc. Nat. Acad. Sci., \textbf{106}, 8847 (2009). 
\bibitem{raval} A. Raval, Phys. Rev. E \textbf{68}, 066119 (2003).  
\bibitem{stehle} J. Stehle, A. Barrat, G. Bianconi, arXiv:1002.4109
\bibitem{watts_sw_98} D.~J. Watts and S.~H. Strogatz, Nature  \textbf{393}, 440 (1998).
\bibitem{latora_efficient_01} V.\ Latora and M.\ Marchiori, Phys. Rev. Lett. \textbf{87}, 198701 (2001).
\bibitem{delosrios} C. Caretta Cartozo and P. De Los Rios
  Phys. Rev. Lett. \textbf{102}, 238703 (2009).
\bibitem{vicsek} A.-L. Barab\'asi, H. Jeong, R. Ravasz, Z. Neda, T. Vicsek, and A. Schubert, Physica  A {\bf 311}, 590 (2002).
\bibitem{clauset_persistence_2007} A.~Clauset and N.~Eagle.  \newblock
  {Persistence and Periodicity in a Dynamic Proximity Network}.
  \newblock In {\em {Proc. of {DIMACS} Workshop on Computational
      Methods for Dynamic Interaction Network}}, 2007.
\bibitem{kempe_connectivity_2002} D.~Kempe, J.~Kleinberg, and A.~Kumar, 
J. Comp. Sys. Sci.,  \textbf{64}, 820 (2002).
\bibitem{kostakos_temporal_2009} V.~Kostakos, Physica A \textbf{388}, 1007 (2009). 
\bibitem{park} J. Park,  M. E. J. Newman, Phys. Rev. E \textbf{70}, 066117 (2004) 
\bibitem{bianconi} G. Bianconi Phys. Rev. E. \textbf{79}, 036114 (2009).  
\bibitem{note} In our discussion, we have implicitly assumed that the 
typical time, $\tau_m$,  to pass a message from a node to one of its 
first neighbors, is of the same order as the typical time, $\tau_g$,  at which the graphs in
the sequence are changing.
We can simulate the case $\tau_m < \tau_g$ by increasing the reach of a message (within a graph in the sequence) past its first neighbors.  Note however that as the reach increases, the values of L will decrease while C does not change; therefore our main results still hold.
\bibitem{infocom} J. Scott, R. Gass, J. Crowcroft, P. Hui, C. Diot, 
and A. Chaintreau, CRAWDAD Trace, Infocom2006. 
\bibitem{facebook} C. Wilson, B. Boe, A. Sala, K.P.N. Puttaswamy, and
  B.Y.Zhao, Procs. of EuroSys '09, pp. 205--218 (2009)

\end{thebibliography}
\end{document}